# Generalized Damping Torque Analysis of Ultra-Low Frequency Oscillation in the Jerk Space

Yichen Zhou, Yang Yang, Tao Zhou, and Yonggang Li

***Abstract*—Ultra low frequency oscillation (ULFO) is significantly threatening the power system stability. Its unstable mechanism is mostly studied via generalized damping torque analysis method (GDTA). However, the analysis still adopts the framework established for low frequency oscillation. Hence, this letter proposes a GDTA approach in the jerk space for ULFO. A multi-information variable is constructed to transform the system into a new state space, where it is found that the jerk dynamics of the turbine-generator cascaded system is a second-order differential equation. Benefiting from this characteristic, we propose a new form for GDTA using jerk dynamics, which is established in the frequency-frequency acceleration phase space. Then, analytical expressions of all damping torque are provided. Finally, test results verified the proposed theoretical results. The negative damping mechanism is revealed, and parameter adjustment measures are concluded.**

## I. Introduction

Ultra-Low frequency oscillation (ULFO) below 0.1Hz is threatening power system stability, especially in high hydro-dominant power systems, e.g. 2011's Colombian power grid ULFO[1], 2016's Yunnan power grid ULFO [2]. It can lead to non-fault tripping and islanding of generators [1].

Understanding the oscillation mechanism is crucial for suppressing ULFO. The common method is the generalized damping torque analysis (GDTA) [2], [3]. However, the old framework of low frequency oscillation is still employed, which emphasizes on rotor angle dynamics. Therefore, a new GDTA framework is needed to investigate ULFO of system frequency.

This letter proposes a novel GDTA method to reveal the ULFO mechanism. Contributions of this letter include: 1) A new framework in the jerk space is established. 2) The frequency - frequency acceleration phase space is proposed for ULFO analysis. 3) Analytical expressions are given for all damping torque in the new framework. 4) Negative damping mechanism and parameter adjustment measures are presented.

## II. Analysis Model of ULFO

As usually occurs in hydro-dominant power systems, the hydropower single-machine single-load system (HSMSL) [2] is often adopted to analyze ULFO. As shown in Fig. 1, we further split the hydraulic turbine model into two forms

$$G_{ht}(s) = (1 - T_w s)(1 + 0.5T_W s)^{-1} = 3(1 + 0.5T_W s)^{-1} - 2 . \quad (1)$$

with their outputs defined as $\Delta T_{md}$ and $\Delta T_{mh}$, respectively.

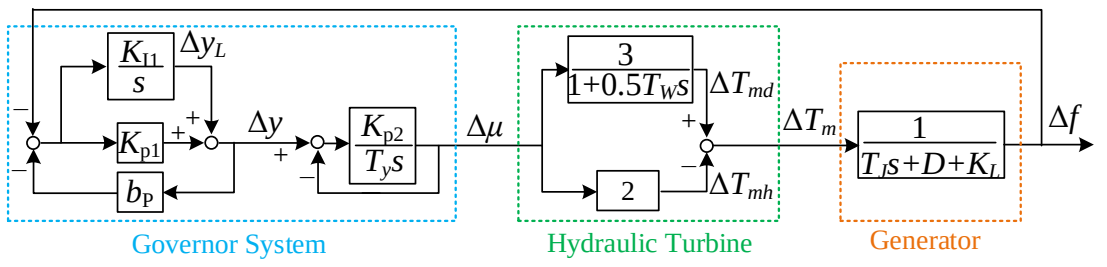

Fig.1 Primary frequency regulation of HSMSL system.

Originally, Fig.1 can be modeled in the state-space as

$$\begin{bmatrix} \Delta\dot{f} \\ \Delta\dot{T}_{md} \\ \Delta\dot{\mu} \\ \Delta\dot{y}_L \end{bmatrix} = \begin{bmatrix} -a_{11} & a_{12} & -a_{13} & 0 \\ 0 & -a_{22} & a_{23} & 0 \\ -a_{31} & 0 & -a_{33} & a_{34} \\ a_{41} & 0 & 0 & -a_{44} \end{bmatrix} \begin{bmatrix} \Delta f \\ \Delta T_{md} \\ \Delta\mu \\ \Delta y_L \end{bmatrix} \quad (2)$$

where $a_{11} = (D + K_L)T_J^{-1}$ , $a_{12} = T_J^{-1}$ , $a_{13} = 2T_J^{-1}$ , $a_{22} = 2T_W^{-1}$ , $a_{33} = K_{P2}T_y^{-1}$ , $a_{23} = 6T_W^{-1}$ , $a_{31} = K_{P2}K_{p1}T_y^{-1}(1 + K_{p1}b_P)^{-1}$ , $a_{34} = K_{P2}T_y^{-1}(1 + K_{p1}b_P)^{-1}$ , $a_{41} = K_{I1}\left(-1 + b_P K_{p1}(1 + K_{p1}b_P)^{-1}\right)$ , $a_{44} = K_{I1}b_P(1 + K_{p1}b_P)^{-1}$ .

To facilitate the ULFO analysis, a multi-information variable $\Delta x = -T_J a_{11}\Delta f + T_J a_{12}\Delta T_{md} - T_J a_{13}\Delta\mu$ is introduced to replace $\Delta T_{md}$ . Then by using the transformation

$$\begin{bmatrix} \Delta f \\ \Delta x \\ \Delta\mu \\ \Delta y_L \end{bmatrix} = \begin{bmatrix} 1 & 0 & 0 & 0 \\ -T_J a_{11} & T_J a_{12} & -T_J a_{13} & 0 \\ 0 & 0 & 1 & 0 \\ 0 & 0 & 0 & 1 \end{bmatrix} \begin{bmatrix} \Delta f \\ \Delta T_{md} \\ \Delta\mu \\ \Delta y_L \end{bmatrix} , \quad (3)$$

the system (2) can be mapped to a new state space express as

$$\begin{bmatrix} \Delta\dot{f} \\ \Delta\dot{x} \\ \Delta\dot{\mu} \\ \Delta\dot{y}_L \end{bmatrix} = \begin{bmatrix} 0 & a_{12} & 0 & 0 \\ b_{21} & -b_{22} & b_{23} & -b_{24} \\ -a_{31} & 0 & -a_{33} & a_{34} \\ a_{41} & 0 & 0 & -a_{44} \end{bmatrix} \begin{bmatrix} \Delta f \\ \Delta x \\ \Delta\mu \\ \Delta y_L \end{bmatrix} \quad (4)$$

with $b_{21} = T_J(a_{13}a_{31} - a_{22}a_{11})$, $b_{22} = a_{11} + a_{22}$ , $b_{24} = T_J a_{13}a_{34}$ , $b_{23} = T_J(a_{12}a_{23} + a_{13}a_{33} - a_{22}a_{13})$ .

## III. Analysis Method in New Phase Space

Based on the state-space model (4), this section will propose the GDTA in the jerk space combined with a new phase space.

Fig.2 shows the HSMSL's block diagram based on (4). The red dashed box outlines the turbine-generator cascaded system with input $\Delta T$ and output $\Delta f$ . The blue dashed box outlines the governor system with input $\Delta f$ and output $\Delta\mu$ & $\Delta y_L$ .

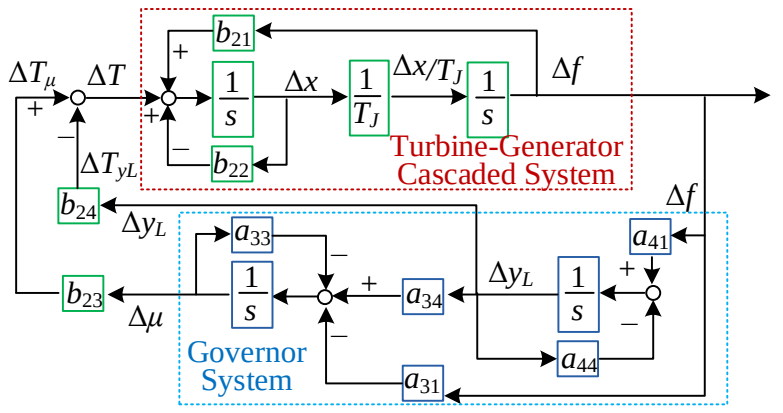

Fig. 2 Diagram for the HSMSL system using the state-space model (4).

Dynamics of the turbine-generator cascaded system (the red dashed box) can be described as

$$T_J\Delta\ddot{f} + T_J b_{22}\Delta\dot{f} - b_{21}\Delta f = \Delta T(s) \quad (5)$$

where total generalized torque $\Delta T$ contains two parts: $\Delta T_{yL}$ and $\Delta T_\mu$ provided by governor system outputs ( $\Delta y_L$ and $\Delta\mu$ ).

Since (5) involves high-order derivatives of $\Delta f$ , we define the first derivative as frequency acceleration $\Delta\alpha = \Delta\dot{f}$ and the second derivative as frequency jerk $\Delta\zeta = \Delta\ddot{f}$. As far as authors know, jerk is introduced for the first time as a degree of freedom in GDTA. Using these definitions, (5) can be rewritten as

$$T_J\Delta\ddot{f} + T_J b_{22}\Delta\alpha - b_{21}\Delta f = \Delta T(s) \quad (6)$$

According to the knowledge of the second-order dynamic system, the oscillation damping of (6) is provided by the term related to $\Delta\alpha$, and the oscillation frequency is affected by the term related to $\Delta f$. To demonstrate above characteristics, a new $\Delta f - \Delta\alpha$ phase space (Fig. 3) is established.

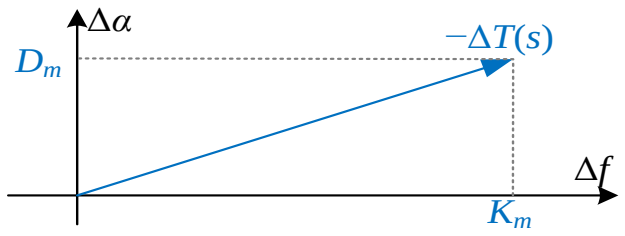

Fig. 3 The $\Delta f - \Delta\alpha$ phase space.

By moving $\Delta T(s)$ to the left of (6), it can be found that the projection of $-\Delta T(s)$ on $\Delta\alpha$ and $\Delta f$ provide generalized damping torque $D_m$ and synchronizing torque $K_m$, respectively.

Based on Fig. 2, let's assume $\Delta T(s) = K(s)\Delta f$. By using the correspondence between $s$-domain and frequency domain, $\Delta T(s)$ will be equal to $\mathrm{Re}(K(s))\Delta f + \mathrm{Im}(K(s))/\omega\,\Delta\alpha$. Hence, the projections are $D_m = -\mathrm{Im}(K(s))/\omega$ and $K_m = -\mathrm{Re}(K(s))$. After adding $D_m$, the total damping becomes

$$D_{total} = T_J b_{22} + D_m \quad (7)$$

which also generates the stability criterion $D_{total} > 0$. And under the influence of $\Delta T$, the expression of ULFO mode is

$$\lambda = -D_{total}/2T_J \pm j\sqrt{D_{total}^2 - 4T_J(b_{21} + K_m)}\Big/2T_J \, . \quad (8)$$

## IV. Damping Analysis of ULFO

This section will analyze all damping sources for ULFO. As exhibited in (7), the first damping term is

$$T_J b_{22} = T_J\left(a_{11} + a_{22}\right) = D + K_L + 2T_J T_W^{-1} \quad (9)$$

which reveals the first three positive damping sources, namely the generator damping coefficient $D$, the load frequency sensitivity $K_L$, and the ratio of generator inertia time constant $T_J$ and water flow inertia time constant $T_W$.

The second damping term $D_m$, provided by the feedback loops, can be divided into two parts as follows

$$D_m = -\mathrm{Im}(K(s))/\omega = \mathrm{Im}(K_{yL}(s))/\omega - \mathrm{Im}(K_\mu(s))/\omega \quad (10)$$

with $K_{yL}(s) = \Delta T_{yL}(s)/\Delta f$ and $K_\mu(s) = \Delta T_\mu(s)/\Delta f$.

Let $a_{31} = a_{34} = 0$, then $\mathrm{Im}(K_{yL})/\omega$ is equal to

$$\frac{-2}{\omega}\mathrm{Im}\left(K_{I1}K_{P2}\left(T_y((1+K_{p1}b_P)s + K_{I1}b_P)(1+K_{p1}b_P)\right)^{-1}\right)\Bigg|_{s=j\omega} \approx 2K_{I1}K_{P2}T_y^{-1}\left|j\omega\left(1+K_{p1}b_P\right) + K_{I1}b_P\right|^{-2} \quad (11)$$

which is positive, serving as the 4th positive damping source.

Let $b_{24} = 0$, then $-\mathrm{Im}(K_\mu(s))/\omega$ is equal to

$$\frac{-1}{\omega}\mathrm{Im}\left(\left(K_{yL} - \frac{2K_{P2}K_{p1}}{T_y(1+K_{p1}b_P)}\right)\frac{T_W^{-1} + K_{P2}T_y^{-1}}{s + K_{P2}T_y^{-1}}\right)\Bigg|_{s=j\omega} \approx \frac{K_{P2}}{T_y}\frac{T_W^{-1} + K_{P2}T_y^{-1}}{\left|j\omega + K_{P2}T_y^{-1}\right|^2}\left(\frac{\mathrm{Im}(K_{yL})}{\omega} + \frac{2\left(b_P K_{I1}^2 + K_{p1}(1+K_{p1}b_P)\omega^2\right)}{\left|(1+K_{p1}b_P)^2\omega^2 + (K_{I1}b_P)^2\right|^2}\right) \quad (12)$$

which is negative, serving as the negative damping source.

Due to the opposite directions of (11) and (13) provided by the governor system, it is essential to clarify their final effect. Therefore, the proportional criterion (13) is proposed as follows

$$\frac{\mathrm{Im}(K_\mu(s))/\omega}{\mathrm{Im}(K_{yL}(s))/\omega} = (T_W^{-1} + K_{P2}T_y^{-1})\left|j\omega + K_{P2}T_y^{-1}\right|^{-2} \left\{K_{P2}T_y^{-1} + K_{I1}[b_P + K_{p1}K_{I1}^{-2}\omega^2(1+K_{p1}b_P)]\right\} \quad (13)$$

Since (9) is positive, the risk of negative damping torque only exists in the governor system. When (13) is large than 1, the governor system will provide negative damping torque.

## V. Case Studies

This section will verify the accuracy of the proposed analysis method and reveal the instability mechanism for ULFO. Typical parameters for the HSMSL test system are shown in Table I. Substituting parameters into (2) and applying eigenvalue analysis will yield a weakly damped ULFO mode at 0.077Hz (-0.0031 ± $j$0.4846).

TABLE I
Typical Parameters of HSMSL System

| $T_J$ | $D$ | $K_L$ | $T_W$ | $K_{P2}$ | $T_y$ | $K_{P1}$ | $K_{I1}$ | $b_P$ |
|---|---|---|---|---|---|---|---|---|
| 6.4 | 1 | 0.03 | 1.5 | 3.36 | 0.2 | 1.7 | 1.2 | 0.04 |

### A. Accuracy Verification

This part employs the relationship (8) between GDTA and eigenvalue analysis to validate the analytical expression of damping torque.

For GDTA, results of (9), (11) and (12) are $T_J b_{22} = 9.56$, $\mathrm{Im}(K_{yL})/\omega = 149.4$ and $\mathrm{Im}(K_\mu)/\omega = 158.92$. Thus, according to (7), the total damping torque $D_{total}$ is 0.04. Then utilizing the relationship (8), the corresponding real part of the ULFO mode should be -0.0031, which is consistent with the above results of eigenvalue analysis. Therefore, the accuracy of the derived damping analysis formulas has been validated, which are applicable for investigating the mechanism of ULFO instability.

### B. Instability Mechanism

As the first part (9) of the total damping torque (7) provides positive damping torque, this part will explore the ULFO instability mechanism triggered by the second part, namely, the damping torque provided by the governor system (10).

By calculating (10) for 500 random system parameter sets, we obtain the results in Fig. 4. It shows that the governor system provides negative damping torque in all scenarios. This phenomenon is reasonable and can be explained by (13). Considering governor servomotor performance and ULFO characteristics, we usually have $T_y \ll 1$, $T_W > 1$, $\omega < 0.63$ and a sufficiently enough $K_{P2}$. Hence, in (13), the first polynomial term, approximately equal to $K_{P2}T_y^{-1}$, is larger than 1, and the second term is larger than 0. Finally, (13) is definitely greater than 1. Therefore, excessive negative damping torque supplied by the governor system is the reason for ULFO instability.

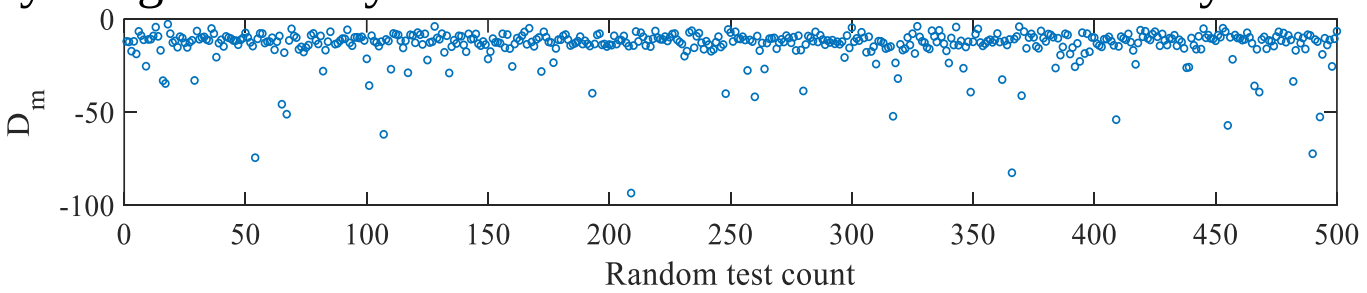

Fig. 4 Damping torque offered by the governor system at 500 random scenarios.

Further analysis is conducted in the frequency jerk space to elucidate the mechanism of the aforementioned negative damping torque. Based on system information (4), the damping torque paths (11) and (12) can be illustrated in both the physical system and the state space block diagram (Fig. 5). Path 1 refers to (11) offered by the integral component of the governor control system. And Path 2 corresponds to (12) supplied by the cascaded governor control and servo system, where Branch 2-1 and 2-2 relate to the first and the second term, respectively.

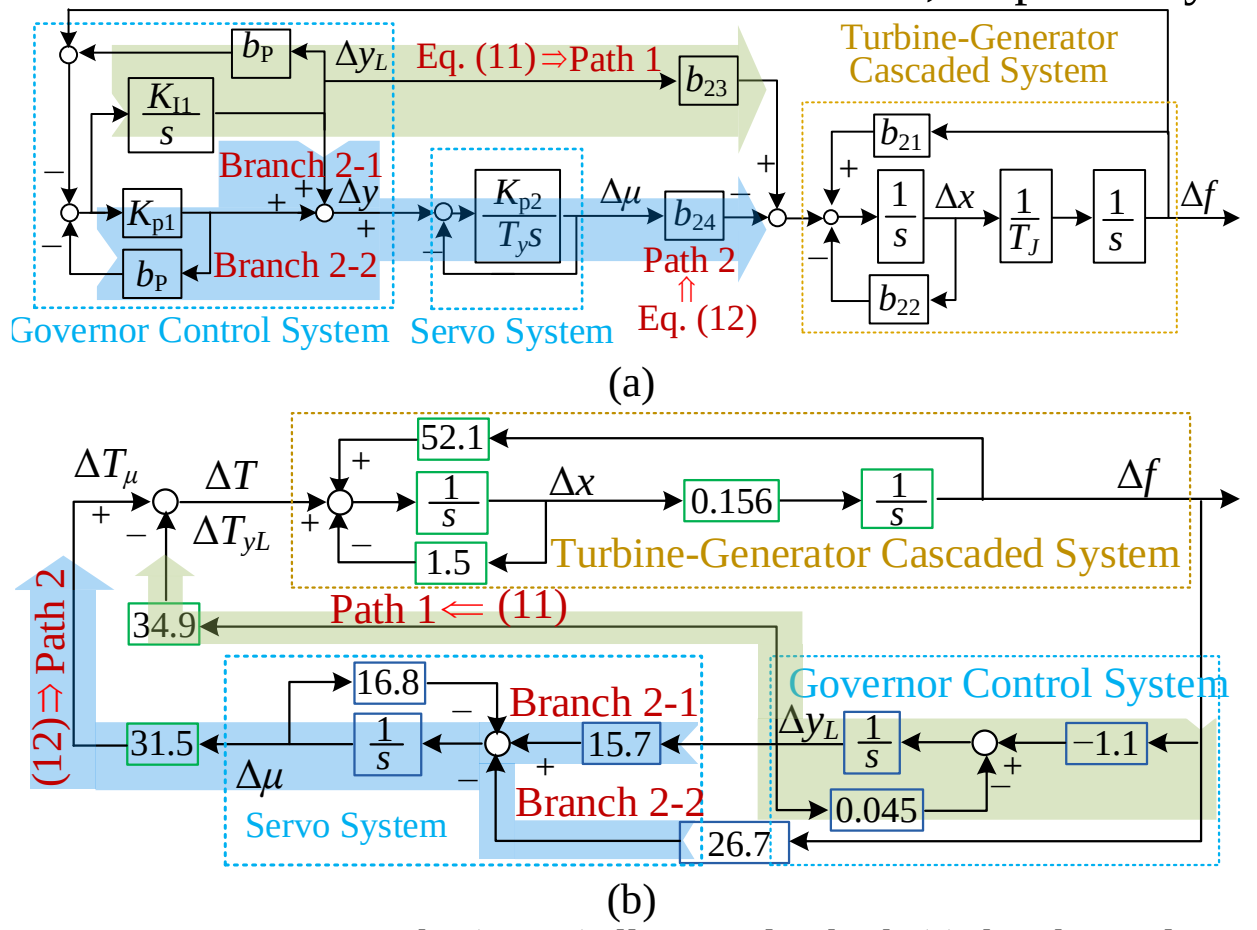

Fig. 5 Damping torque paths (11-12) illustrated in both (a) the physical system diagram and (b) the state space block diagram with parameters in Table I.

Usually, $a_{33}$ is relative large and $a_{44}$ is relative small under the influence of small $T_y$ and $b_P$. To make mechanism analysis clear, typical parameters in Table I are taken as an example to draw Fig. 5(b), but the conclusions also apply to other parameters. The instability mechanism is shown in Fig. 6. Assuming a small disturbance on $\Delta x$, governor system outputs through Path 1 and two branches of Path 2 are approximately $\Delta y_L \approx -C_1/s^2\,\Delta x$, $\Delta\mu \approx -C_4/s^2\,\Delta f$ and $\Delta\mu \approx -C_6/s^2\,\Delta x$, respectively. Substituting $s = j\omega$, we can obtain that all governor outputs are increased under disturbance. Further considering $\Delta\dot{x} \propto -\Delta T_{yL}$ and $\Delta T_{\mu}$, the positive $\Delta T_{yL}$ from Path 1 will suppress the disturbance. However, the positive $\Delta T_{\mu}$ from two branches of Path 2 will amplify the disturbance. Therefore, the main reason for ULFO instability is the excessive cancellation resulting from the damping competition between the integral loop of the governor (Path 1) and the governor itself (Path 2).

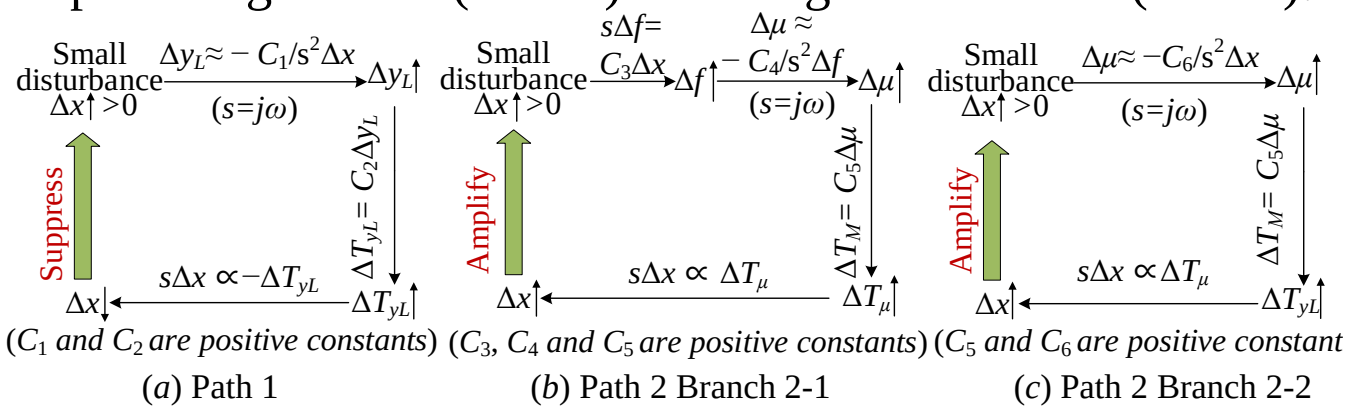

Fig. 6 Instability mechanism for ULFO.

## C. Parameter Adjustment for Stability Enhancement

Adjusting system parameters without adding equipment is the most cost-effective measure to enhance ULFO stability.

Numerically, the ratio of Branch 2-1's damping torque to Path 1's damping torque is $\left(T_W^{-1}+K_{P2}T_y^{-1}\right)K_{P2}T_y^{-1}\left|j\omega+K_{P2}T_y^{-1}\right|^{-2}$, larger than 1. Therefore, reducing this ratio and decreasing Path 1's damping torque will both contribute to reducing Branch 2-1's negative damping torque. According to (11), Path 1's damping torque can be reduced by decreasing $K_{I1}$ or increasing $b_P$, whose role in enhancing stability is verified by Fig. 7(a), (b), and (e). When $K_{P1}$ is small, the ratio can be reduced by increasing $K_{P1}$, thus the stability will be improved as illustrated by Fig.7(a), (c), (e) and Fig.8.

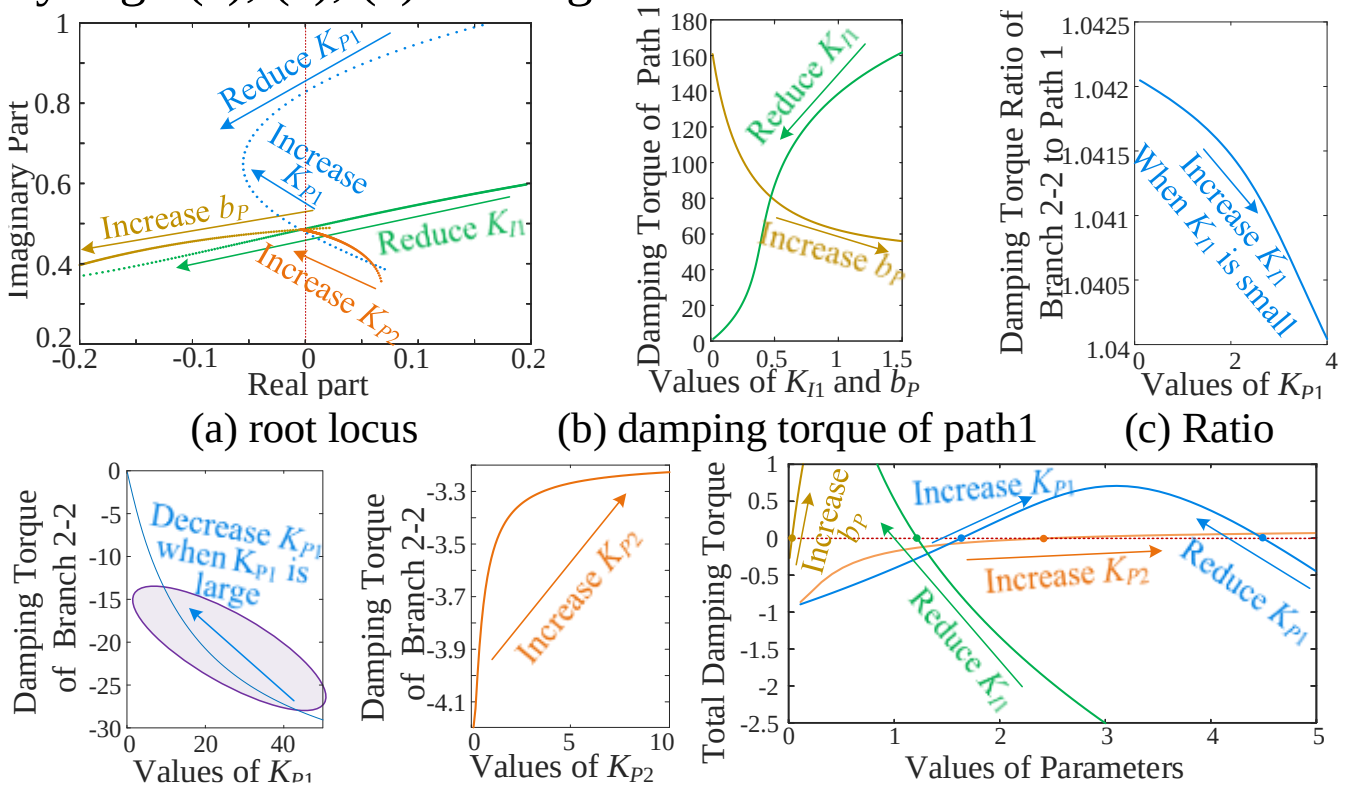

(a) root locus (b) damping torque of path1 (c) Ratio

(d)Damping torque of branch 2-2 (e) Total Damping torque

Fig. 7 Effects of Parameter adjustment.

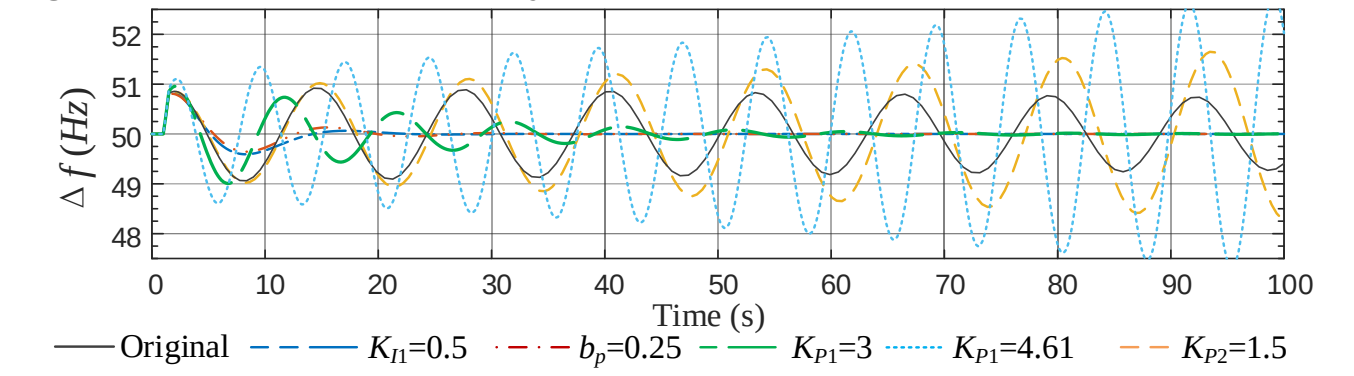

Fig. 8 Simulink/Matlab simulation results for parameter adjustment measures.

In addition to reducing the total negative damping torque generated by both Path 1 and Branch 2, stability can also be enhanced by decreasing the negative damping torque of Branch 2-2. According to (12), by increasing $K_{P2}$ or decreasing $K_{P1}$ when $K_{P1}$ is large, the negative damping torque of branch 2-2 can be reduced, which is proved by Fig.7 (a), (d), (e) and Fig.8.

Besides, it can be inferred from positive damping torque (9) that the stability can also be improved by adjusting the system operating state to obtain a larger $D$, $T_J$, $K_L$ and a smaller $T_W$.

## VI. Conclusions

This letter analyzed the ULFO by using the proposed GDTA of jerk space. Significant conclusions are: 1) It is verified by both the phenomenon and theoretical explanation that the governor system always provides negative damping torque. 2) The mechanism of ULFO instability is the competition between the positive damping torque related to the governor's integral loop and the negative damping torque related to the whole governor, with the positive damping torque being at a disadvantage. 3) Parameter adjustment measures to improve the ULFO stability include reducing negative damping torque of Branch 2-1 ($K_{I1}\downarrow$, $b_P\uparrow$) and Branch 2-2 ($K_{P2}\uparrow$), and increasing positive damping torque related to system characteristics ($D\uparrow$, $T_J\uparrow$, $K_L\uparrow$, $T_W\downarrow$).